\documentclass[letter]{aa}

\usepackage{verbatim}
\usepackage{graphicx}
\usepackage{epstopdf}
\usepackage{subfigure}
\usepackage{amsmath}
\usepackage{hyperref}
\usepackage{tabularx}
\usepackage{longtable}

\begin{document} 

\title{Extreme hydrodynamic losses of Earth-like atmospheres in the habitable zones of very active stars}

\titlerunning{Extreme losses from Earth-like atmospheres}

\author{C. P. Johnstone\inst{\ref{vienna}} \and M. L. Khodachenko\inst{\ref{graz},\ref{moscow},\ref{siberia}} \and T. L\"uftinger\inst{\ref{vienna}} \and K. G. Kislyakova\inst{\ref{vienna},\ref{graz}} \and H. Lammer\inst{\ref{graz}} \and M. G\"udel\inst{\ref{vienna}}}
\institute{
University of Vienna, Department of Astrophysics, T\"{u}rkenschanzstrasse 17, 1180 Vienna, Austria \label{vienna}
\and
Space Research Institute, Austrian Academy of Sciences, Graz, Austria \label{graz}
\and
Skobeltsyn Institute of Nuclear Physics, Moscow State University, Moscow, Russia \label{moscow}
\and
Institute of Astronomy of the Russian Academy of Sciences, 119017, Moscow, Russia  \label{siberia}
}

\abstract{}{
In this letter, we calculate for the first time the full transonic hydrodynamic escape of an Earth-like atmosphere.
We consider the case of an Earth-mass planet with an atmospheric composition identical to that of the current Earth orbiting at 1 AU around a young and very active solar mass star. 
}{
To model the upper atmosphere, we used the Kompot Code, which is a first-principles model that calculates the physical structures of the upper atmospheres of planets, taking into account hydrodynamics and the main chemical and thermal processes taking place in the upper atmosphere of a planet. 
This model enabled us to calculate the 1D vertical structure of the atmosphere using as input the high-energy spectrum of a young and active Sun.
}{
The atmosphere has the form of a transonic hydrodynamic Parker wind, which has an outflow velocity at the upper boundary of our computational domain that exceeds the escape velocity. 
The outflowing gas is dominated by atomic nitrogen and oxygen and their ion equivalents and has a maximum ionization fraction of 20\%.
The mass outflow rate is found to be \mbox{$1.8 \times 10^9$~g~s$^{-1}$}, which would erode the modern Earth's atmosphere in less than 0.1~Myr.
}{
This extreme mass loss rate suggests that an Earth-like atmosphere cannot form when the planet is orbiting within the habitable zone of a very active star.
Instead, such an atmosphere can only form after the activity of the star has decreased to a much lower level.
This happened in the early atmosphere of the Earth, which was likely dominated by other gases such as CO$_2$.
Since the time it takes for the activity of a star to decay is highly dependent on its mass, this is important for understanding possible formation timescales for planets orbiting low-mass stars. 
}

\maketitle


\section{Introduction} \label{sect:intro}

Planetary atmospheres lose gas to space largely as a consequence of interactions between the upper atmosphere and the central star.
Most important is the X-ray and ultraviolet (XUV) radiation of the star, used in this work to mean the wavelength range  1--400~nm.
Since it is absorbed at high altitudes where the gas densities are low, XUV radiation is able to heat the upper atmosphere to very high temperatures, potentially causing the gas to flow away from the planet hydrodynamically. 
The evolution of the atmosphere of a planet is therefore closely linked to the evolution of the XUV radiation of the star.

The XUV emission of a star depends on its magnetic field, which is a function of the mass, rotation rate, and age of the star. 
Rapidly rotating stars emit more XUV than slowly rotating stars (\citealt{Pizzolato03}), and since stars spin down with age, their XUV radiation also decreases (\citealt{Guedel97}).
Recently, \citet{Tu15} showed that the XUV evolution of a star depends sensitively on its initial rotation rate and that evolutionary tracks that are very different in the first Gyr are possible.
\citet{Johnstone15letter} showed that this can be very important for the evolution of the atmosphere of a planet.
To complicate matters, the activity lifetimes of lower mass stars, and especially M dwarfs, are significantly longer than those of higher mass stars (\citealt{West08}).

For the solar system planets, the main physical processes taking place in the upper atmospheres of planets are broadly well understood (e.g., \citealt{Roble88}; \citealt{Fox91}). 
The absorption of XUV radiation in the upper atmosphere drives a large number of chemical and thermal processes, creating a region of high temperatures called the thermosphere, and a region of partially ionized gas called the ionosphere.
The results of this heating and ionization is an enhancement of atmospheric losses to space by a large number of mechanisms, including for example ion outflows from the magnetic poles (\citealt{Glocer09}) and exospheric pick-up by stellar winds (\citealt{Kislyakova14}).

Using hydrostatic models, \citet{Kulikov07} studied the upper atmospheres of the Earth assuming a very active Sun and found thermospheric temperatures exceeding 20,000~K for their most active cases. 
\cite{Tian08} studied the response of the upper atmosphere of the Earth to higher solar XUV activity levels using a more complete physical model that also included hydrodynamic effects. 
They found that for high XUV fluxes, very high Jeans escape at the exobase causes a significant upward flow of material leading to adiabatic cooling of the gas.
Since these authors did not consider cases in which the Sun was very active, the atmospheres that they modeled were not transonic hydrodynamic flows that exceed the escape velocity at the exobase.
\citet{Lichtenegger10} used the atmospheres from \cite{Tian08} to calculate the losses due to ionization and pick up of exospheric gas by the solar wind and estimated mass loss rates on the order $10^7$~g~s$^{-1}$, which would remove the atmosphere of the Earth in 10~Myr. 

Fully hydrodynamic models of planetary upper atmospheres have so far only been applied to H and He dominated atmospheres (e.g., \mbox{\citealt{Khodachenko15}}; \mbox{\citealt{OwenMohanty16}}; \mbox{\citealt{Shaikhislamov18}}) and to an atomic H and O gas by \citet{Guo19}.
In this letter, we study for the first time the hydrodynamic outflow of an atmosphere composed of heavier molecules orbiting a very active star. 
We study whether such atmospheres are hydrodynamic, how rapidly mass is lost, and what such an atmosphere looks like in terms of physical structure.
In Section~\ref{sect:model} we describe our model, in Section~\ref{sect:results} we present the results, and in Section~\ref{sect:conclusions} we discuss our conclusions.

\begin{figure}[!t]
\centering
\includegraphics[trim = 0mm 2mm 0mm 2mm, clip=true,width=0.49\textwidth]{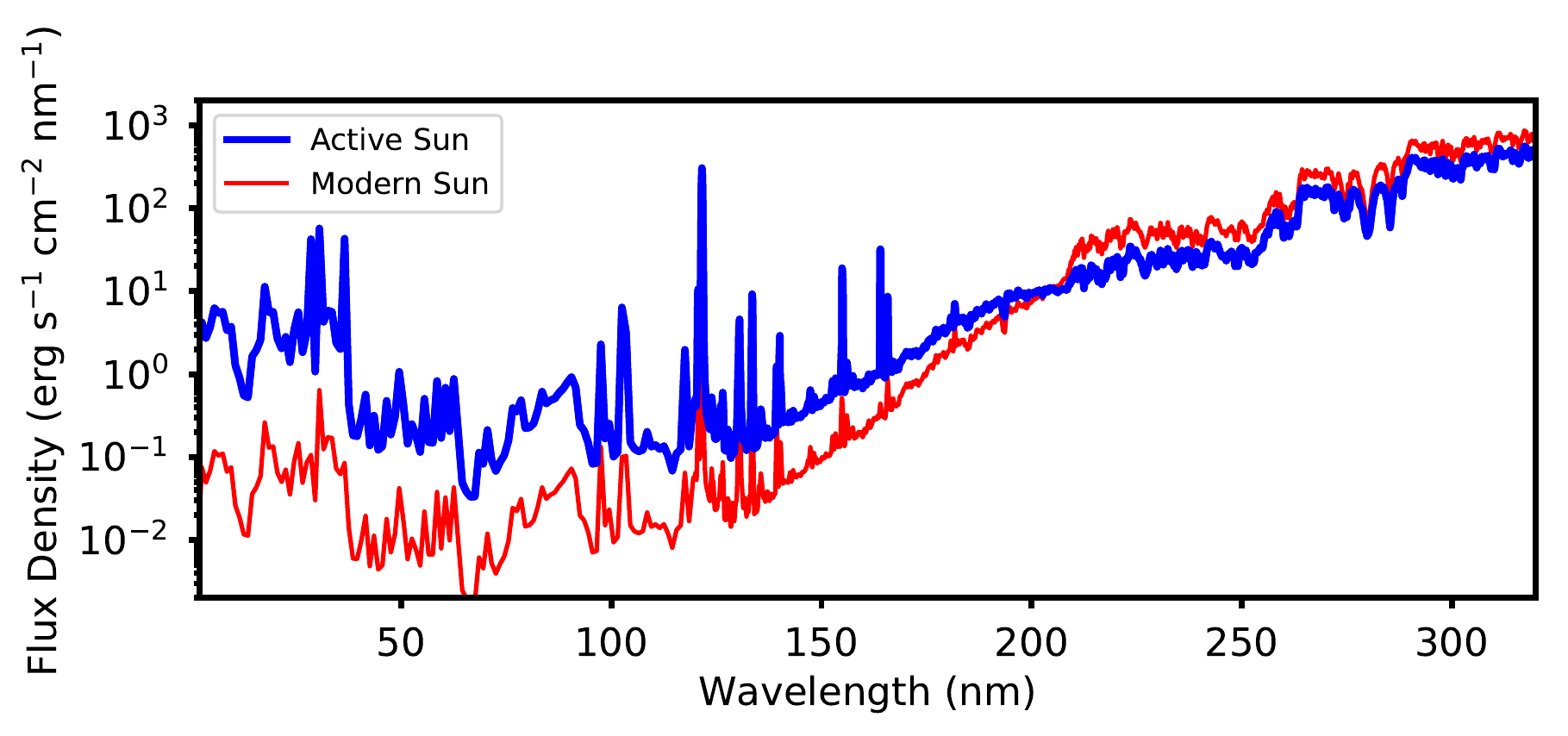}
\includegraphics[trim = 0mm 2mm 0mm 2mm, clip=true,width=0.49\textwidth]{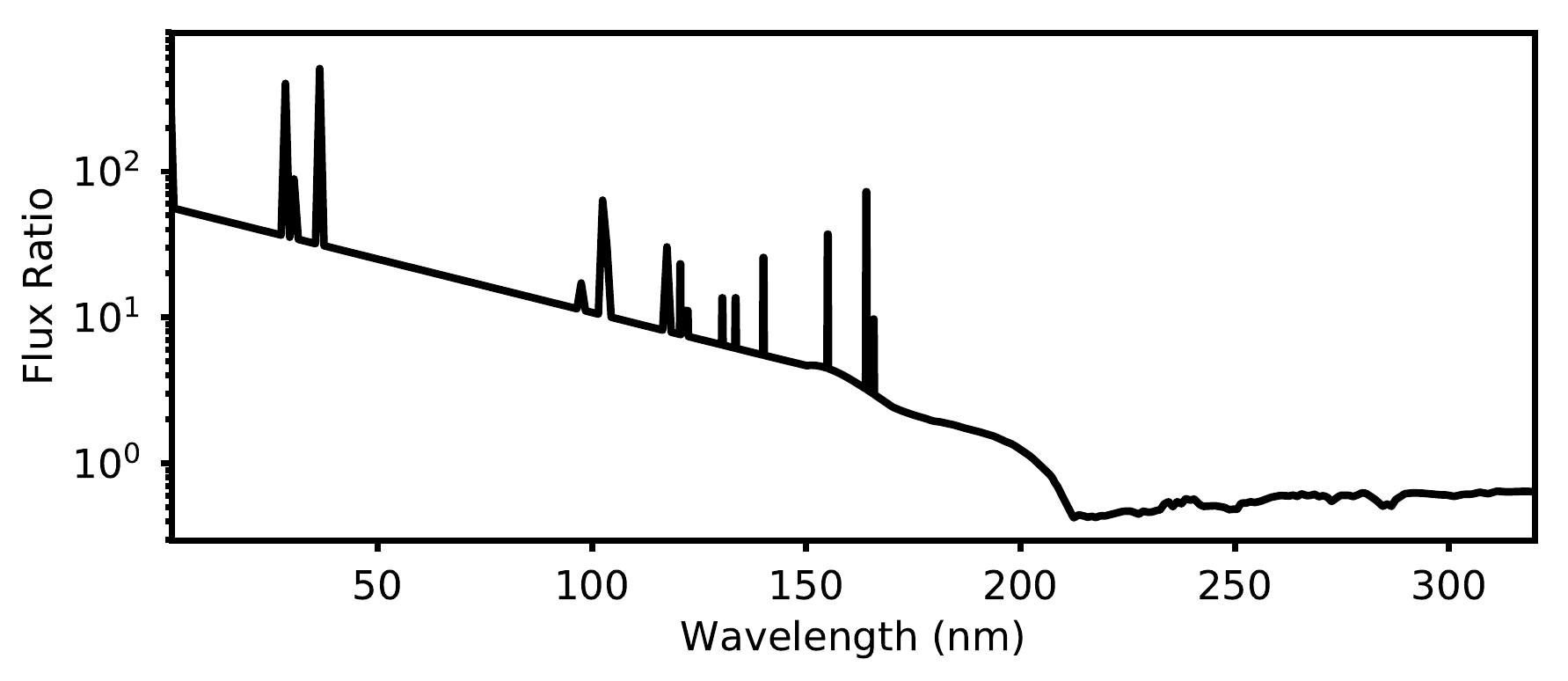}
\caption{
Typical modern solar activity maximum spectrum and the XUV spectrum used in our model as calculated by \citet{Claire12}.
The upper panel shows the two spectra and the lower panel shows the ratio of the two.
\vspace{-5mm}
}
\label{fig:spectra}
\end{figure}

\begin{figure}[!h]
\centering
\includegraphics[trim = 0mm 0mm 0mm 0mm, clip=true,width=0.45\textwidth]{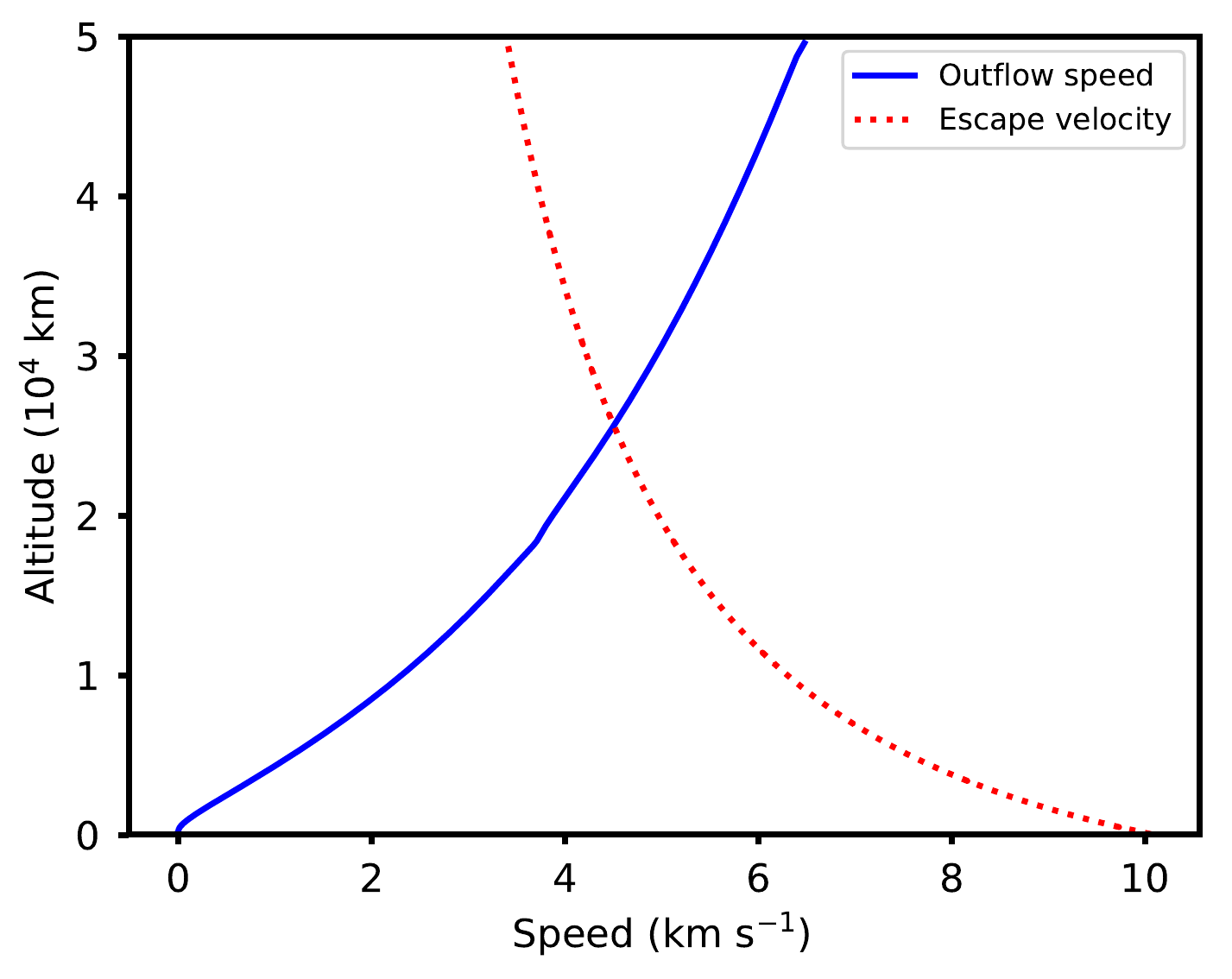}
\includegraphics[trim = 0mm 0mm 0mm 0mm, clip=true,width=0.45\textwidth]{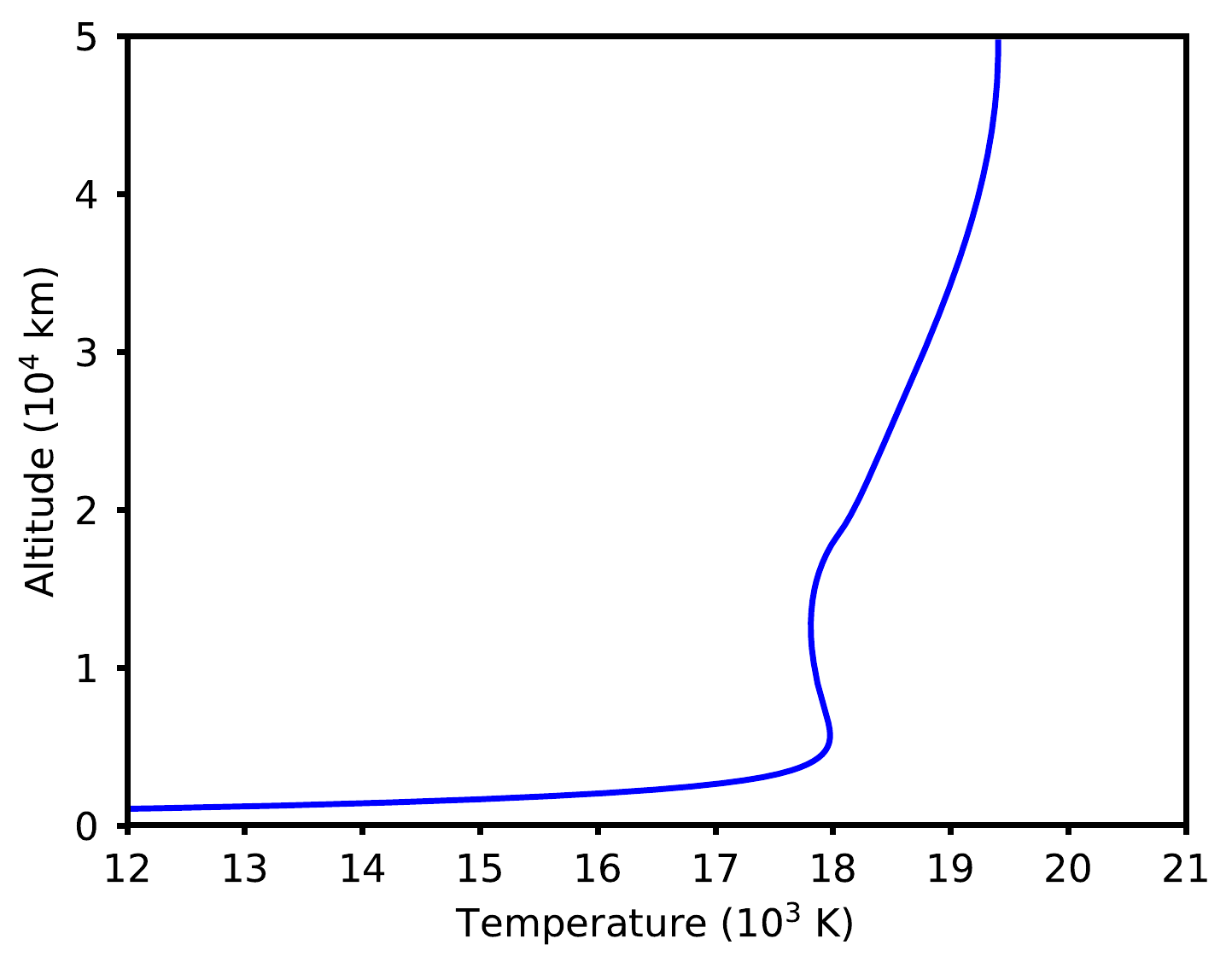}
\includegraphics[trim = 0mm 0mm 0mm 0mm, clip=true,width=0.45\textwidth]{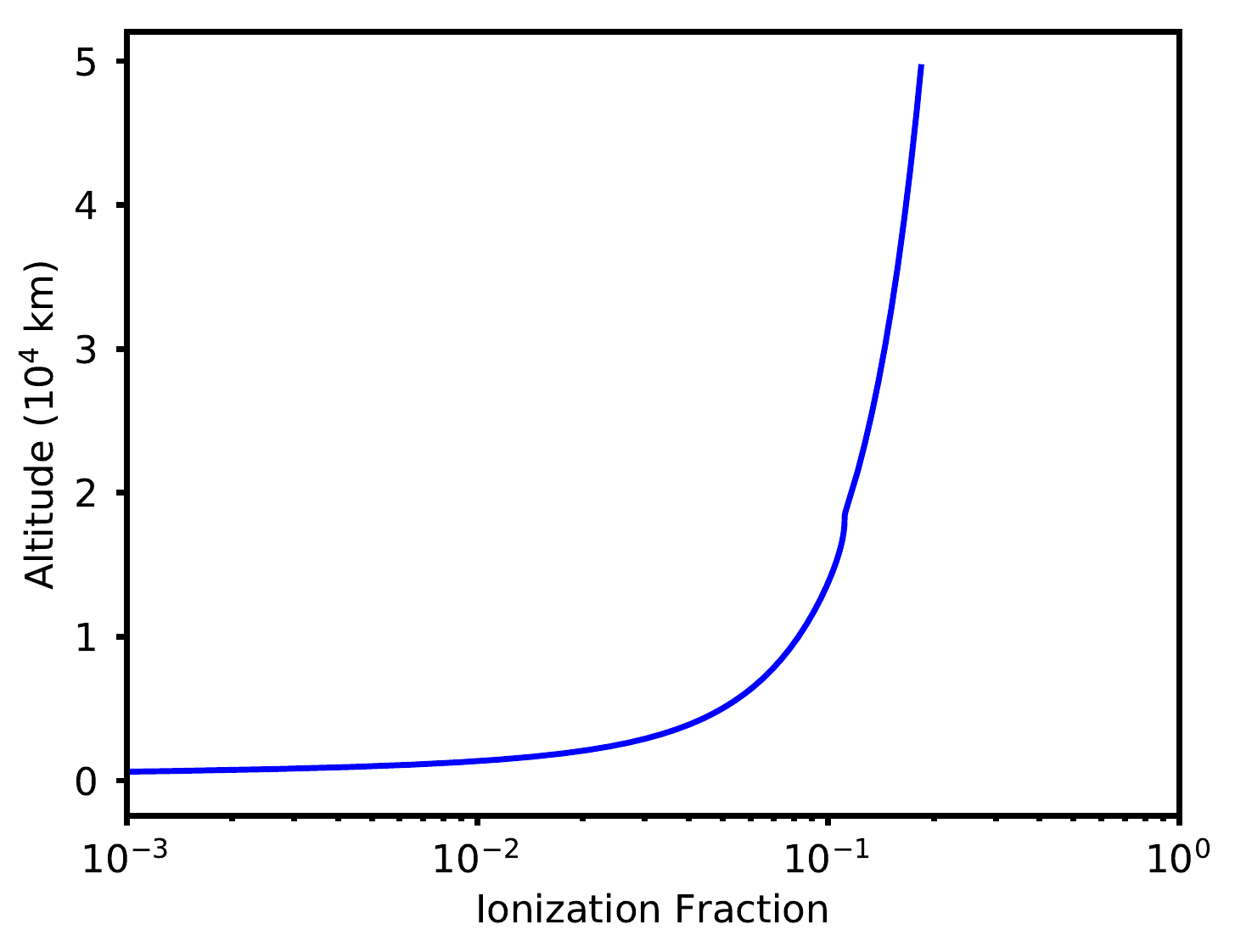}
\caption{
Outflow speed, temperature, and ionization fraction of the gas in our simulation.
\vspace{-5mm}
}
\label{fig:hydroresults}
\end{figure}

\section{Model} \label{sect:model}


We consider the case of an Earth mass and radius planet orbiting a solar mass star at 1~AU.
The 1D computational domain extends between altitudes of 50 and 50,000~km and we assume a zenith angle of 66$^\circ$.
The domain is broken down into 500 cells with the radial cell width increasing linearly with radius by a factor of 1000 between the lower and upper boundaries.
At the lower boundary, the temperature and species densities are held constant at the values typical for the Earth at 50~km altitude used in \citet{Johnstone18} and the advection speed is held at zero. 
At the upper boundary, zero-gradient outflow conditions are used.

For the solar XUV spectrum between 1 nm and 400~nm, we used the spectrum of a very active Sun as estimated by \citet{Claire12}.
Specifically we used their estimate of the solar spectrum from 4.4~Gyr ago, although we stress that the solar spectrum at this age would have actually depended sensitively on the rotation rate of the Sun (\citealt{Tu15}).
The X-ray luminosity of the star in this case is \mbox{$10^{29}$~erg~s$^{-1}$}, which is approximately the 50th percentile of the distribution of X-ray luminosities at 100~Myr for solar mass stars. 
The input spectrum, a typical modern solar spectrum, and the ratio of the two are shown in Fig.~\ref{fig:spectra}.
Relative to the modern spectrum, the flux in the X-ray and extreme ultraviolet (EUV) part of the input spectrum ($\lambda<100$~nm) is enhanced by a factor of 60, while the flux in the entire spectrum is reduced by 30\%. 
Our input spectrum is significantly more active than the most active case considered by \citet{Tian08}, which had the EUV enhanced by a factor of 20. 

To calculate the atmospheric structure we used the Kompot Code, recently developed by \citet{Johnstone18}.
The model is designed to include, as much as has so far been possible, all of the most important physical processes implemented based on first-principles physical considerations. 
See \citet{Johnstone18} for detailed descriptions of the model (Section~2) and the numerical solvers (appendices).
The thermal processes we considered are heating by stellar XUV radiation, cooling by the emission of infrared radiation to space, thermal conduction, and hydrodynamic transport. 
The XUV heating model includes several components, including heating by exothermic chemical reactions and high-energy photoelectrons; this model has the advantage that it does not use any unconstrained free parameters such as the commonly used heating efficiency.
To calculate the chemical structure of the atmosphere, we considered chemistry, including XUV photochemistry, diffusion, and hydrodynamic advection.
For the chemistry, we considered 503 reactions, including 56 photoreactions, and we considered a total of 63 species, of which 30 are ions. 
We solved the chemistry in a time-dependent way, meaning no equilibrium assumption was made and we evolved the abundances of each chemical species in time with all other quantities in the model (see Appendix~H of \citealt{Johnstone18}).
For diffusion, we considered both eddy and molecular diffusion.
The eddy diffusion coefficients are the only real free parameters in the model and we simply assumed the values from the modern Earth case, which is likely realistic since our atmosphere profile is similar to that of the modern Earth below 100~km where eddy diffusion is important. 

We solved the full set of hydrodynamic equations to model the outflow of the atmosphere.
We replace the numerical scheme described in \citet{Johnstone18} with the MUSCL-Hancock scheme described in Section~14.6.3 of Toro~(1999).
The major difference with this scheme is that it performs the timestep update on the primitive variables (mass density, velocity, and thermal pressure) instead of on the conservative variables (mass, momentum, and energy densities).
We find that for atmospheric hydrodynamic purposes, primitive variable schemes are more reliable.

The major simplification that we make to the model of \citet{Johnstone18} is we remove the assumption that the neutrals, ions, and electrons have separate temperatures. 
This has been done primarily because the Riemann solver used, described in Section~9.3 (Eqn.~9.28) of Toro~(1999), is only appropriate for single temperature gases.
Based on the simulations of \citet{Johnstone18} for the Earth under higher solar XUV fluxes (see their Fig.~13), we would expect the three temperatures to be similar in any case.
In the single temperature model, all of the various heating and cooling mechanisms are included. 
The only processes that we exclude entirely are the energy exchanges between the neutrals, ions, and electrons, which are not required.
For the thermal conductivity, we calculated separate values for the neutrals, ions, and electrons, as described in \citet{Johnstone18}, and then used the density weighted average of these three\footnotemark.
Eddy conduction is included.

\footnotetext{
Taking the density weighted average thermal conductivity is a simplification and is the procedure recommended in Section~22.7.2 of \citet{BanksKockarts73} for ion mixtures.
A more detailed calculation requires knowledge of the collisional cross sections between each component (e.g., \citealt{Orrall61}).  
Tests have shown that changes of more than an order of magnitude are necessary to influence our results significantly.
Our conductivities vary from those of purely neutral and fully ionized gases by much smaller amounts. 
}

\begin{figure}
\centering
\includegraphics[trim = 0mm 0mm 0mm 0mm, clip=true,width=0.45\textwidth]{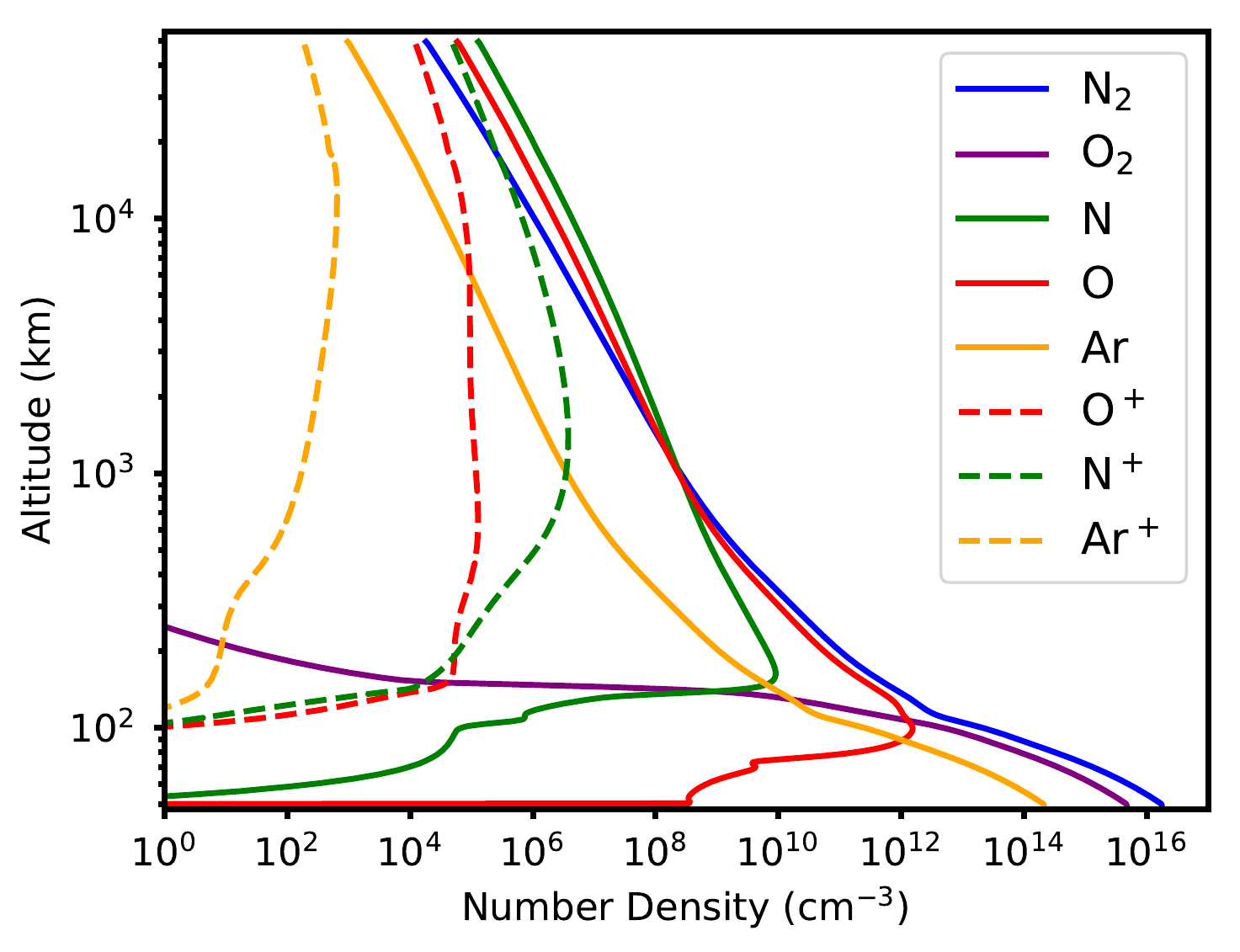}
\includegraphics[trim = 0mm 0mm 0mm 0mm, clip=true,width=0.45\textwidth]{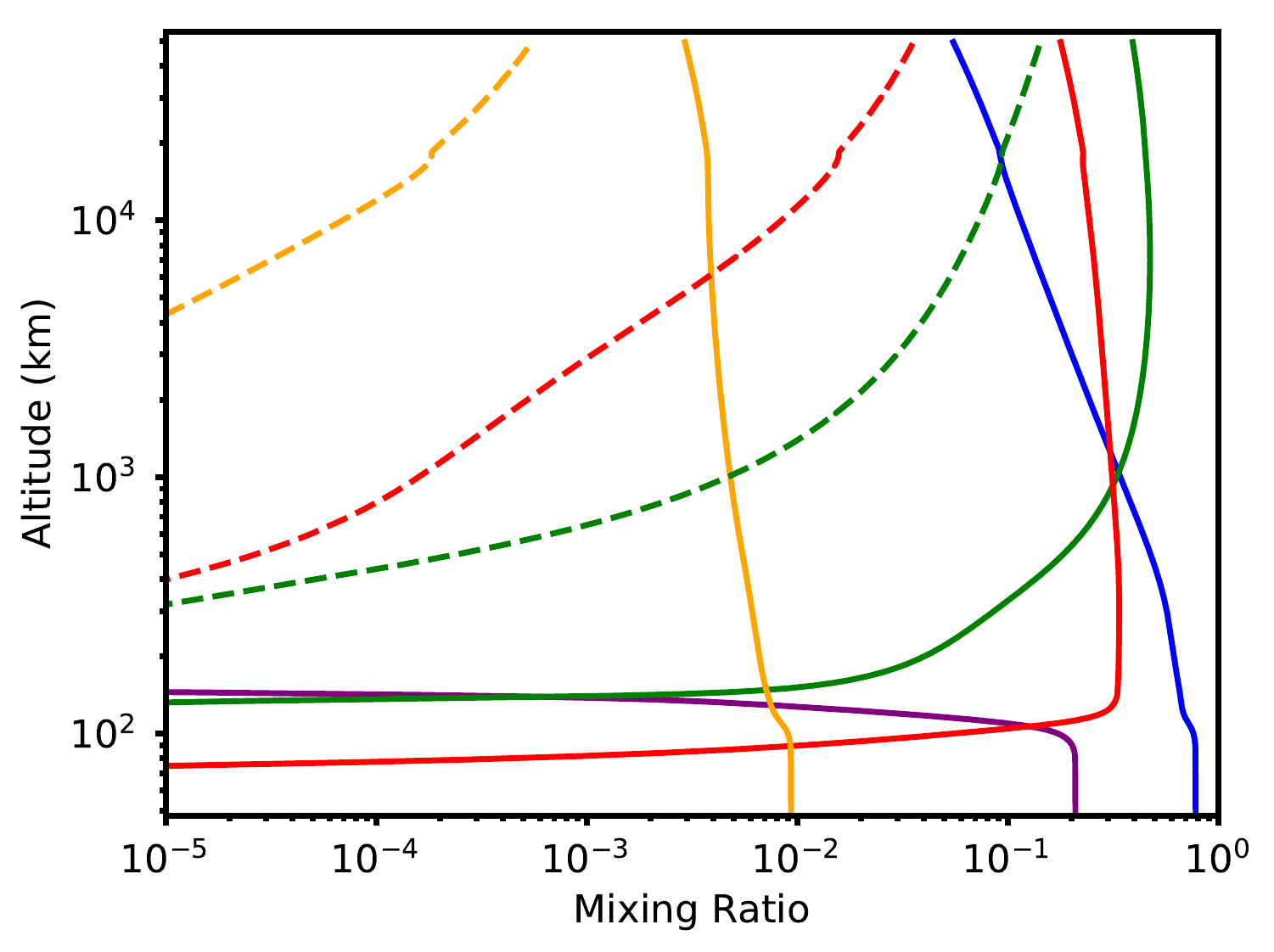}
\caption{
Densities and mixing ratios of 8 of the species in the simulation, as listed in the upper panel.
The mixing ratio at each location is defined as the number density of the species divided by the total number density of the gas. 
\vspace{-4mm}
}
\label{fig:chemresults}
\end{figure}

\section{Results} \label{sect:results}

We find that under the very large XUV flux of the young Sun, the upper atmosphere of the Earth is heated to such a high temperature that it flows away from the planet in the form of a transonic hydrodynamic wind, which is similar in structure to the hydrodynamic outflows that have been modeled for H/He-dominated atmospheres.
Vertical profiles for the outflow speed, temperature, and ionization fraction (defined as the fraction of particles that are ions) are shown in Fig.~\ref{fig:hydroresults}.
The exobase is not within the boundaries of the computational domain and would likely extend beyond the boundaries of a magnetosphere with a size similar to that of the Earth.  
At the upper boundary of the simulation, the temperature is 19,000~K and the outflow speed is 7~km~s$^{-1}$ with an escape velocity of 4~km~s$^{-1}$, meaning that the gas flows freely away from the planet.
Our atmosphere is hotter than that of the highest solar activity case considered by \citet{Tian08} because our case has a much higher input XUV flux.
We also do not see decreasing temperatures at increasing altitude; although adiabatic cooling is an important effect, at such high XUV fluxes, the XUV driven heating is stronger than adiabatic cooling. 

The chemical structure of the atmosphere is shown in Fig.~\ref{fig:chemresults}; the upper and lower panels show the densities and mixing ratios of several species, respectively.
The three species shown that are stable in the lower atmosphere are N$_2$, O$_2$, and Ar.
The mixing ratios of these species are uniform below the homopause at approximately 100~km; this region is the homosphere and it is caused by convective mixing, implemented in our model as eddy diffusion.
Above the homopause, XUV photochemistry and molecular diffusion, which separates the species by mass, cause a rapid change in the chemical composition of the gas. 
Between 100 km and 300~km, O$_2$ is rapidly photodissociated and atomic oxygen becomes the second most abundant species.
Above 1000~km, photodissociation of N$_2$ causes N to become the most abundant species, and free electrons become the second most abundant species by the top of the computational domain.
The chemical composition above approximately 2000~km changes very slowly owing to the importance of hydrodynamic advection, which has the effect (similar to that of eddy diffusion) of making the chemical composition uniform.
The only significant changes to the composition above 2000~km are due to photoionization, where N$^+$ and O$^+$ become the most abundant ions by the upper boundary. 

\begin{figure}
\centering
\includegraphics[trim = 0mm 0mm 0mm 0mm, clip=true,width=0.45\textwidth]{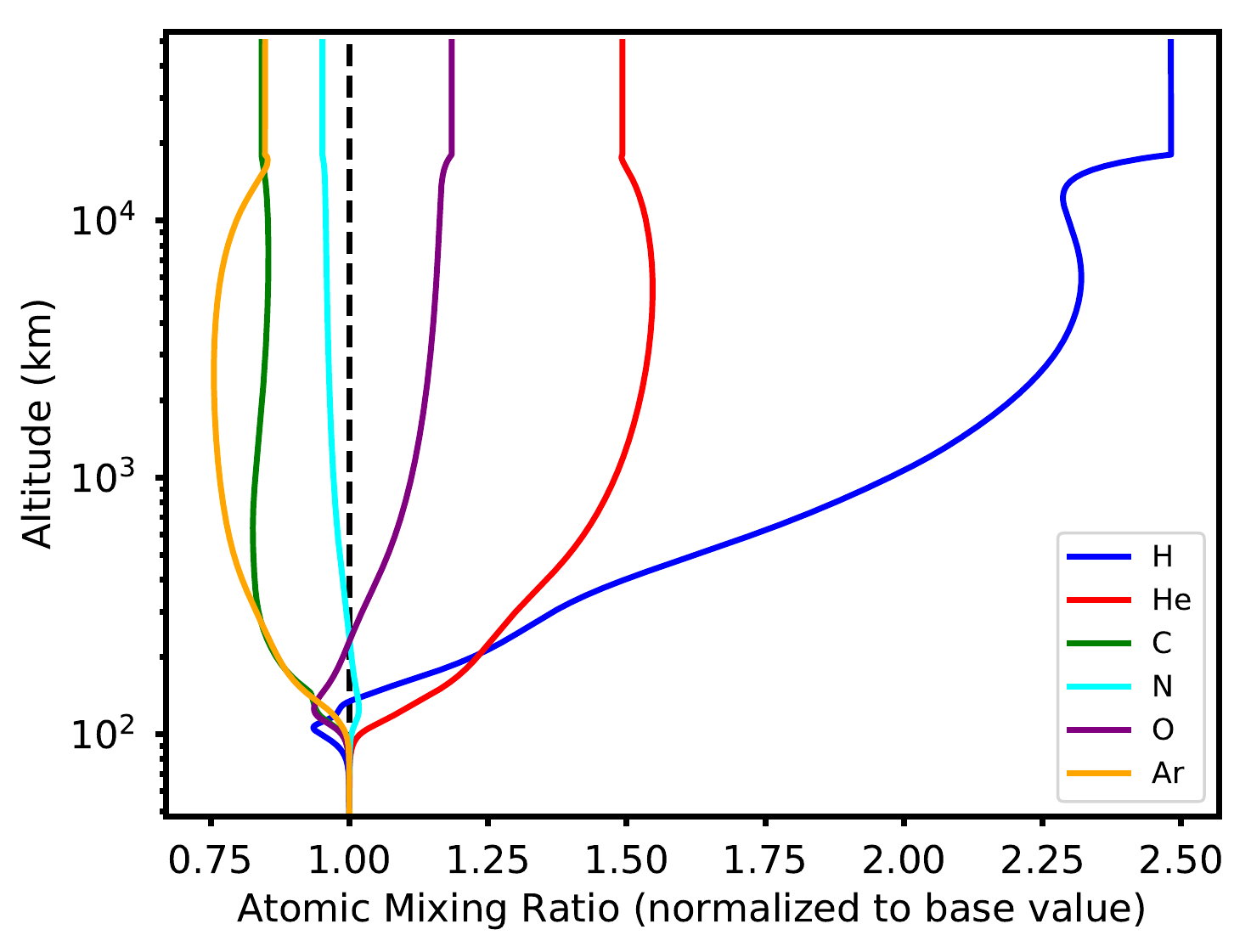}
\includegraphics[trim = 0mm 0mm 0mm 0mm, clip=true,width=0.45\textwidth]{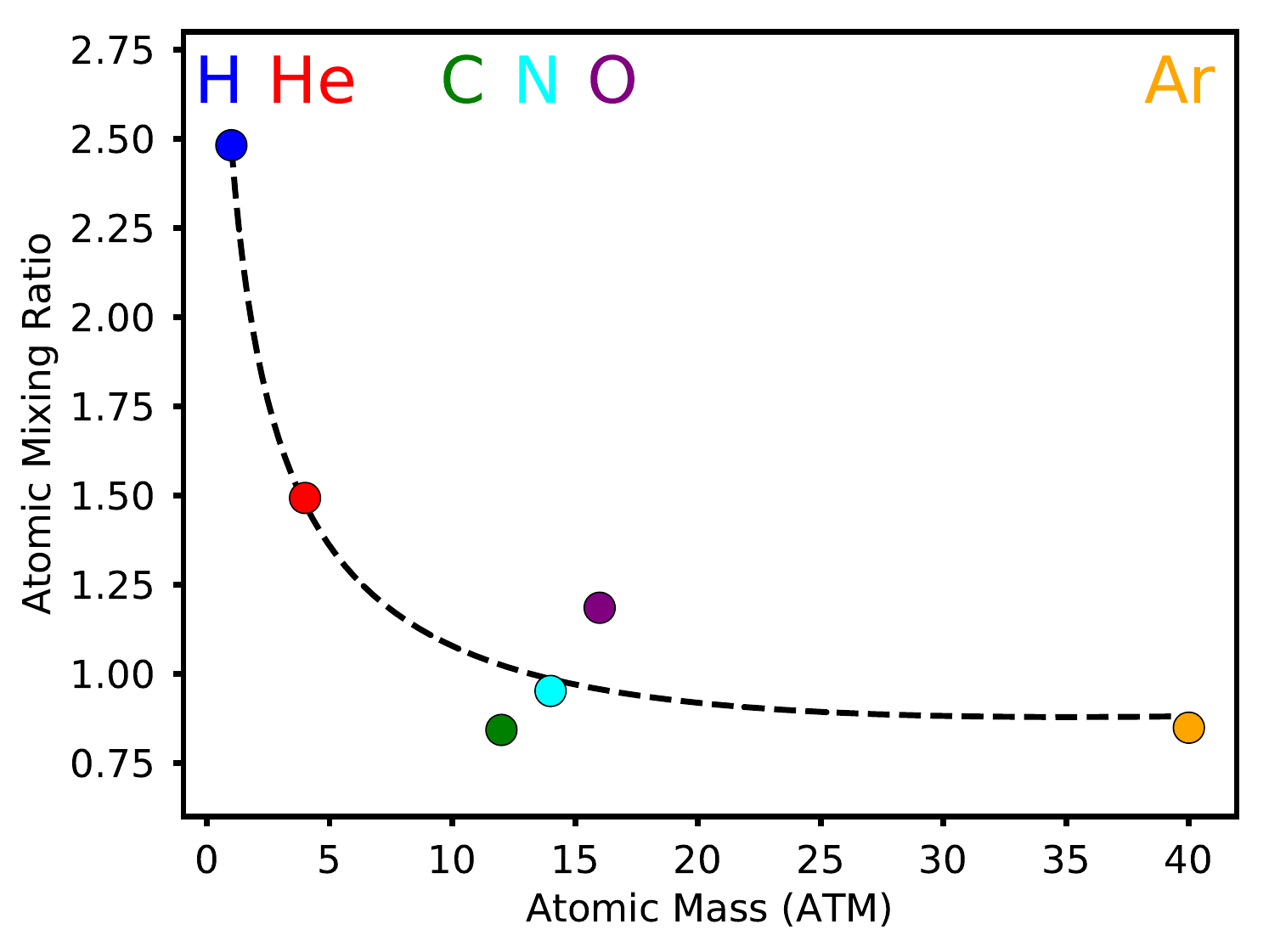}
\caption{
Atomic mixing ratios (see text for a definition) of individual elements as functions of altitude (\emph{upper panel}), and mixing ratios at the upper boundaries of the simulation for each element as a function of elemental mass (\emph{lower panel}).
All atomic mixing ratio values are normalized to the values at the lower boundary of the simulation. 
The lines show some unrealistic behavior in the upper regions where the atmosphere becomes supersonic; for numerical efficiency and stability, we only calculate diffusion in the subsonic part of the atmosphere.
\vspace{-4mm}
}
\label{fig:elementresults}
\end{figure}

It is interesting to consider not only the chemical composition of the gas, but also the total elemental composition.
The six elements in our simulation are H, He, C, N, O, and Ar.
In Fig.~\ref{fig:elementresults}, we show the atomic mixing ratios of each element as functions of altitude and the values at the upper boundary of our simulation as a function of atomic mass.
The atomic mixing ratio is defined at each altitude as the total density of atoms (including those contained in molecules) of a given element divided by the total density of atoms.
By dividing all values in Fig.~\ref{fig:elementresults} by the values at the base of the simulation, we get a measure of how efficiently individual species are lost relative to each other. 
In general, lighter elements are lost more efficiently than heavier elements owing to molecular diffusion above the homopause separating the species by mass; H is lost $\sim$2.5 times more efficiently than most other species, suggesting that water vapor (the source of H in our simulation) is lost more rapidly than other important atmospheric species.
Mass is however not the only factor, which can be seen from the fact that N is lost less effectively than the heavier O. This is likely because O$_2$ is more rapidly photodissociated than N$_2$ in the lower thermosphere. 
This difference is however small; N and O are lost at approximately the same rate relative to their abundances in the lower atmosphere.
Even though this element is even less massive, C is lost even less efficiently, most likely because in the lower atmosphere it is contained within heavy CO$_2$ molecules.

The mass loss rate in the simulation is \mbox{$1.8 \times 10^{9}$~g~s$^{-1}$}.
As a comparison, the commonly used energy-limited formula for hydrodynamic loss is \mbox{$\dot{M} = \epsilon \pi F_\mathrm{xuv} R_\mathrm{pl} R_\mathrm{xuv}^2 / G M_\mathrm{pl}$}, where $\epsilon$, $F_\mathrm{xuv}$, and $R_\mathrm{xuv}$ are the mass loss efficiency, input XUV flux, and radius at which XUV radiation is absorbed (\citealt{Luger15}; \citealt{ChenRogers16}). 
If we calculate $F_\mathrm{xuv}$ as the total flux in the spectrum at the top of the simulation minus the value at the lower boundary, we get \mbox{$F_\mathrm{xuv}=855$~erg~s$^{-1}$~cm$^{-2}$}.
\textrm{We note that this is an order-of-magnitude estimate for the appropriate $F_\mathrm{xuv}$ since our simulation domain is at a zenith angle of 66$^\circ$ and the radiation is not traveling directly downward.}
The reason this is so high is that it also includes radiation absorbed in the UV spectrum between 100 and 300~nm.
Assuming \mbox{$\epsilon=1$} and \mbox{$R_\mathrm{xuv}=R_\mathrm{pl}$}, we get a mass loss rate of \mbox{$1.7 \times 10^{9}$~g~s$^{-1}$}, which is almost identical to our modeled value. 
The fact that these values are almost identical is partly coincidental and in reality \mbox{$\epsilon<1$} and \mbox{$R_\mathrm{xuv}>R_\mathrm{pl}$}, but the comparison shows that our very high mass loss rate is reasonable. 
If we just consider X-ray and EUV radiation, the energy-limited formula gives a mass loss rate of \mbox{$5.7 \times 10^{8}$~g~s$^{-1}$}, which is a factor of three lower.
This suggests a difference between Earth-like atmospheres and primordial H/He atmospheres.
Since the main components of primordial atmospheres do not absorb XUV radiation effectively at wavelengths longer than $\sim$100~nm, the energy at these wavelengths is absorbed deeper in the atmosphere where the density is higher and the effect on the gas is lower.
Earth-like atmospheres however are composed of many species, particularly molecules, that absorb energy effectively at wavelengths up to $\sim$200~nm, and even effectively up to $\sim$300~nm when O$_3$ is present.

\section{Conclusions} \label{sect:conclusions}



The mass loss rate in our model is \mbox{$1.8 \times 10^{9}$~g~s$^{-1}$,} which would remove the entire atmosphere of the Earth in approximately 0.1~Myr. 
Over evolutionary timescales, this is essentially instantaneous and means that an Earth-like N$_2$ and O$_2$ atmosphere can only form within the habitable zone of a star after its activity has decayed.
It is important to note that the age at which a activity level of  a star decays depends sensitively on its initial rotation rate (\citealt{Johnstone15}; \citealt{Tu15}).
This has important implications for the evolution of Earth-like atmospheres in the habitable zones of low-mass M dwarfs, since such stars have very long activity lifetimes (\citealt{West08}).
It has even been suggested that some M dwarfs do not even spin down significantly (\citealt{Irwin11}), and might therefore always remain highly active, suggesting that Earth-like atmospheres can never form in such systems. 
Further work is needed to test if the extreme mass loss rates are similar for planets orbiting active M dwarfs that have different spectral shapes (\citealt{Fontenla16}), especially at wavelengths above $\sim$150~nm because of their cooler photospheres. 

Several processes could protect an atmosphere from being lost to space during the active phase of the host star.
For the early Earth, we expect that the atmosphere during the most active phase of the Sun was significantly different and that CO$_2$ is the most probable candidate for the dominant component (\citealt{Lammer18}).
As a strong emitter of infrared radiation, CO$_2$ could have cooled the upper atmosphere (\citealt{Kulikov07}; \citealt{Johnstone18}) and reduced the atmospheric losses significantly.
Given the high ionization fractions that we find in the upper thermosphere, it is possible that the intrinsic magnetic field of a planet could also play a role in reducing the mass loss rate, as was shown to be possible for the atmospheres of Hot Jupiters by \citet{Khodachenko15}.
Also, volatile species can only be lost to space if they are indeed in the atmosphere and the volatile content of a planet can be protected from being lost simply by being held within the crust and mantle and then being released after the star's activity has decayed to much lower levels.
Similarly, even if a planet loses its entire atmosphere, it is often possible for a new atmosphere to form from the reservoirs of volatiles buried below the surface, which can be released to form an atmosphere by a range of processes (\citealt{Noack14}). 

It is interesting that the atmosphere in our simulation exceeds the escape velocity below the exobase and therefore escapes hydrodynamically. 
This is the first time that it has been shown that an Earth-like N$_2$ and O$_2$ dominated atmosphere, or any atmosphere composed of molecules heavier than $H_2$, would become fully hydrodynamic within the habitable zone of a very active star.
In future research, we will study these processes in more detail, including developing a fully 3D model for upper atmospheres and studying the effects of planetary rotation on the hydrodynamic outflow.

\section{Acknowledgments} 

We authors thank the referee for useful comments and suggestions on the manuscript of this letter.
This study was carried out with the support by the FWF NFN project S11601-N16 ``Pathways to Habitability: From Disk to Active Stars, Planets and Life'' and the related subprojects S11604-N16, S11606-N16, and S11607-N16. 
We also acknowledge the support by grant 16-52-14006 of the Russian Fund of Basic Research, grant 18-12-00080 of the Russian Science Foundation, and the Austrian Science Foundation (FWF) project I2939-N27.
T.L. acknowledges funding via the Austrian Space Application Programme (ASAP) of the Austrian Research Promotion Agency (FFG) within ASAP11.

\bibliographystyle{aa}
\bibliography{mybib}

\begin{thebibliography}{27}
\expandafter\ifx\csname natexlab\endcsname\relax\def\natexlab#1{#1}\fi

\bibitem[{{Banks} \& {Kockarts}(1973)}]{BanksKockarts73}
{Banks}, P.~M. \& {Kockarts}, G. 1973, {Aeronomy.}

\bibitem[{{Chen} \& {Rogers}(2016)}]{ChenRogers16}
{Chen}, H. \& {Rogers}, L.~A. 2016, \apj, 831, 180

\bibitem[{{Claire} {et~al.}(2012){Claire}, {Sheets}, {Cohen}, {Ribas},
  {Meadows}, \& {Catling}}]{Claire12}
{Claire}, M.~W., {Sheets}, J., {Cohen}, M., {et~al.} 2012, \apj, 757, 95

\bibitem[{{Fontenla} {et~al.}(2016){Fontenla}, {Linsky}, {Witbrod}, {France},
  {Buccino}, {Mauas}, {Vieytes}, \& {Walkowicz}}]{Fontenla16}
{Fontenla}, J.~M., {Linsky}, J.~L., {Witbrod}, J., {et~al.} 2016, \apj, 830,
  154

\bibitem[{{Fox} \& {Bougher}(1991)}]{Fox91}
{Fox}, J.~L. \& {Bougher}, S.~W. 1991, \ssr, 55, 357

\bibitem[{{Glocer} {et~al.}(2009){Glocer}, {T{\'o}th}, {Gombosi}, \&
  {Welling}}]{Glocer09}
{Glocer}, A., {T{\'o}th}, G., {Gombosi}, T., \& {Welling}, D. 2009, Journal of
  Geophysical Research (Space Physics), 114, A05216

\bibitem[{{G{\"u}del} {et~al.}(1997){G{\"u}del}, {Guinan}, \&
  {Skinner}}]{Guedel97}
{G{\"u}del}, M., {Guinan}, E.~F., \& {Skinner}, S.~L. 1997, \apj, 483, 947

\bibitem[{{Guo}(2019)}]{Guo19}
{Guo}, J.~H. 2019, \apj, 872, 99

\bibitem[{{Irwin} {et~al.}(2011){Irwin}, {Berta}, {Burke}, {Charbonneau},
  {Nutzman}, {West}, \& {Falco}}]{Irwin11}
{Irwin}, J., {Berta}, Z.~K., {Burke}, C.~J., {et~al.} 2011, \apj, 727, 56

\bibitem[{{Johnstone} {et~al.}(2018){Johnstone}, {G{\"u}del}, {Lammer}, \&
  {Kislyakova}}]{Johnstone18}
{Johnstone}, C.~P., {G{\"u}del}, M., {Lammer}, H., \& {Kislyakova}, K.~G. 2018,
  \aap, 617, A107

\bibitem[{{Johnstone} {et~al.}(2015{\natexlab{a}}){Johnstone}, {G{\"u}del},
  {L{\"u}ftinger}, {Toth}, \& {Brott}}]{Johnstone15}
{Johnstone}, C.~P., {G{\"u}del}, M., {L{\"u}ftinger}, T., {Toth}, G., \&
  {Brott}, I. 2015{\natexlab{a}}, \aap, 577, A27

\bibitem[{{Johnstone} {et~al.}(2015{\natexlab{b}}){Johnstone}, {G{\"u}del},
  {St{\"o}kl}, {Lammer}, {Tu}, {Kislyakova}, {L{\"u}ftinger}, {Odert},
  {Erkaev}, \& {Dorfi}}]{Johnstone15letter}
{Johnstone}, C.~P., {G{\"u}del}, M., {St{\"o}kl}, A., {et~al.}
  2015{\natexlab{b}}, \apjl, 815, L12

\bibitem[{{Khodachenko} {et~al.}(2015){Khodachenko}, {Shaikhislamov}, {Lammer},
  \& {Prokopov}}]{Khodachenko15}
{Khodachenko}, M.~L., {Shaikhislamov}, I.~F., {Lammer}, H., \& {Prokopov},
  P.~A. 2015, \apj, 813, 50

\bibitem[{{Kislyakova} {et~al.}(2014){Kislyakova}, {Johnstone}, {Odert},
  {Erkaev}, {Lammer}, {L{\"u}ftinger}, {Holmstr{\"o}m}, {Khodachenko}, \&
  {G{\"u}del}}]{Kislyakova14}
{Kislyakova}, K.~G., {Johnstone}, C.~P., {Odert}, P., {et~al.} 2014, \aap, 562,
  A116

\bibitem[{{Kulikov} {et~al.}(2007){Kulikov}, {Lammer}, {Lichtenegger}, {Penz},
  {Breuer}, {Spohn}, {Lundin}, \& {Biernat}}]{Kulikov07}
{Kulikov}, Y.~N., {Lammer}, H., {Lichtenegger}, H.~I.~M., {et~al.} 2007, \ssr,
  129, 207

\bibitem[{{Lammer} {et~al.}(2018){Lammer}, {Zerkle}, {Gebauer}, {Tosi},
  {Noack}, {Scherf}, {Pilat-Lohinger}, {G{\"u}del}, {Grenfell}, {Godolt}, \&
  {Nikolaou}}]{Lammer18}
{Lammer}, H., {Zerkle}, A.~L., {Gebauer}, S., {et~al.} 2018, \aapr, 26, 2

\bibitem[{{Lichtenegger} {et~al.}(2010){Lichtenegger}, {Lammer},
  {Grie{\ss}meier}, {Kulikov}, {von Paris}, {Hausleitner}, {Krauss}, \&
  {Rauer}}]{Lichtenegger10}
{Lichtenegger}, H.~I.~M., {Lammer}, H., {Grie{\ss}meier}, J.-M., {et~al.} 2010,
  \icarus, 210, 1

\bibitem[{{Luger} {et~al.}(2015){Luger}, {Barnes}, {Lopez}, {Fortney},
  {Jackson}, \& {Meadows}}]{Luger15}
{Luger}, R., {Barnes}, R., {Lopez}, E., {et~al.} 2015, Astrobiology, 15, 57

\bibitem[{{Noack} {et~al.}(2014){Noack}, {Godolt}, {von Paris}, {Plesa},
  {Stracke}, {Breuer}, \& {Rauer}}]{Noack14}
{Noack}, L., {Godolt}, M., {von Paris}, P., {et~al.} 2014, \planss, 98, 14

\bibitem[{{Orrall} \& {Zirker}(1961)}]{Orrall61}
{Orrall}, F.~Q. \& {Zirker}, J.~B. 1961, \apj, 134, 63

\bibitem[{{Owen} \& {Mohanty}(2016)}]{OwenMohanty16}
{Owen}, J.~E. \& {Mohanty}, S. 2016, \mnras, 459, 4088

\bibitem[{{Pizzolato} {et~al.}(2003){Pizzolato}, {Maggio}, {Micela},
  {Sciortino}, \& {Ventura}}]{Pizzolato03}
{Pizzolato}, N., {Maggio}, A., {Micela}, G., {Sciortino}, S., \& {Ventura}, P.
  2003, \aap, 397, 147

\bibitem[{{Roble} {et~al.}(1988){Roble}, {Ridley}, {Richmond}, \&
  {Dickinson}}]{Roble88}
{Roble}, R.~G., {Ridley}, E.~C., {Richmond}, A.~D., \& {Dickinson}, R.~E. 1988,
  \grl, 15, 1325

\bibitem[{{Shaikhislamov} {et~al.}(2018){Shaikhislamov}, {Khodachenko},
  {Lammer}, {Berezutsky}, {Miroshnichenko}, \& {Rumenskikh}}]{Shaikhislamov18}
{Shaikhislamov}, I.~F., {Khodachenko}, M.~L., {Lammer}, H., {et~al.} 2018,
  \mnras, 481, 5315

\bibitem[{{Tian} {et~al.}(2008){Tian}, {Kasting}, {Liu}, \& {Roble}}]{Tian08}
{Tian}, F., {Kasting}, J.~F., {Liu}, H.-L., \& {Roble}, R.~G. 2008, Journal of
  Geophysical Research (Planets), 113, E05008

\bibitem[{{Tu} {et~al.}(2015){Tu}, {Johnstone}, {G{\"u}del}, \&
  {Lammer}}]{Tu15}
{Tu}, L., {Johnstone}, C.~P., {G{\"u}del}, M., \& {Lammer}, H. 2015, \aap, 577,
  L3

\bibitem[{{West} {et~al.}(2008){West}, {Hawley}, {Bochanski}, {Covey}, {Reid},
  {Dhital}, {Hilton}, \& {Masuda}}]{West08}
{West}, A.~A., {Hawley}, S.~L., {Bochanski}, J.~J., {et~al.} 2008, \aj, 135,
  785

\end{thebibliography}

\end{document}